\documentclass[nonacm]{acmart}

\AtBeginDocument{%
  }

\begin{document}

\title{Towards Attention-Aware Large Language Models: Integrating Real-Time Eye-Tracking and EEG for Adaptive AI Responses}


\author{Dan Zhang}
\affiliation{%
 \institution{University of Texas at Austin}
 \city{Austin}
 \state{Texas}
 \country{USA}}



\begin{abstract}
\noindent\rule{\textwidth}{0.5pt}
\vspace{0.5em}
\noindent\textbf{Abstract}

\noindent This project proposes an attention-aware LLM that integrates EEG and eye-tracking to monitor and measure user attention dynamically. To realize this, the project will integrate real-time EEG and eye-tracking data into an LLM-based interactive system and classify the user’s attention state on the fly. The system can identify five attention states: High Attention, Stable Attention, Dropping Attention, Cognitive Overload, and Distraction. It responds accordingly to each state, with a particular focus on adapting to decreased attention, distraction, and cognitive overload to improve user engagement and reduce cognitive load.
\end{abstract}



\maketitle
\hrule
\vspace{1em}
\section{Introduction}
Consider an interactive AI assistant that adjusts its communication style based on the user’s attentional state. For instance, if the system detects the user’s attention is waning, it can pose engaging questions or highlight key points to recapture focus. Currently, most large language models (LLMs) interact with users in a standardized manner, disregarding variations in attention levels or cognitive load. Such a static interaction model may inadvertently lead to disengagement, cognitive fatigue, or even information overload, especially in contexts involving sustained attention or complex learning tasks. By contrast, human-like, attention-aware LLMs could intelligently adapt the timing, content, and delivery of their responses, optimizing user engagement and minimizing unnecessary distraction by continuously recognizing fluctuations in users’ attentional states. Achieving this advanced level of adaptivity requires a deep understanding of attentional dynamics, reliable sensing of physiological and behavioral indicators, and effective strategies for responsive system design.
\par Prior work has emphasized the importance and benefits of attention-aware systems in enhancing human–computer interaction. Early foundational research by Chen and Vertegaal\cite{chen2004using} introduced Physiologically Attentive User Interfaces (PAUI, which can distinguish four degrees of attentional states: at rest, moving, thinking and busy by combining Heart Rate Variability (HRV) signals and electroencephalogram (EEG) analysis. Their work showed the potential of attention aware system in reducing demands on the user's mental load.  Building on this potential, Bailey and Konstan \cite{bailey2006need} further highlighted that attention-aware systems, by carefully managing interruptions and user attention, could significantly reduce task errors, completion times, and negative affective responses, thus substantially improving user experience. Complementing these studies, Roda and Thomas \cite{roda2006attention} provided a comprehensive review emphasizing three interconnected aspects of attention management: identifying the user's current attentional state, assessing alternative points of focus, and devising effective strategies to present these alternatives. Additionally, their work reinforced the importance of integrating cognitive psychology principles, particularly regarding top-down control of attention and collaborative task environments. Together, these studies establish a robust theoretical and empirical basis for developing effective attention-aware systems.
\par To practically implement these theoretical insights, subsequent research has increasingly turned toward physiological measures, such as EEG, to quantify and track attentional states more directly and precisely. Early efforts in EEG-based attention-aware systems typically focused on detecting limited levels of attentional states. For example, Chen et al. \cite{chen2017assessing} developed an attention-aware system designed to accurately detect students' attention levels as a low or high-attention level using EEG signals. The study demonstrated that this system could help online instructors evaluate students' attentiveness more effectively, ultimately enhancing their online learning outcomes. However, this study classified only two classes of attention, and it is not sufficient as attention has more than two levels. To overcome this limitation, Li et al. \cite{li2011real} designed a real-time EEG-based brain computer interface (BCI) to measure attention level. Eight participants were asked to perform four kinds of mental tasks and they needed to record their mental state after they answering tasks question. Three attention levels were classified using a KNN classifier based on the Self-Assessment Manikin (SAM) model. And the average accuracy rate reached 57.03\,\% after seven session’s EEG training. Mohammadpour and Mozaffari [5] further expanded the EEG-based classification from three to four distinct attention levels. Despite this progress, their approach still lacked the granularity necessary for precise attention tracking, suggesting EEG alone might be inadequate for capturing the full complexity of human attention. 
\par To address the limitations of unimodal EEG-based methods, Srinivasan et al. \cite{srinivasan2024} introduced attention-aware visualizations leveraging multiple modalities, including eye-tracking, head orientation, and pointer movements, to dynamically measure and track user attention over time. These visualizations demonstrated how multimodal integration can offer richer and more nuanced insights into attentional states, aligning with a broader trend in human-computer interaction (HCI) to incorporate cognitive and physiological aspects into system design. These studies highlight the progressive evolution from simple EEG-based classifications to integrated multimodal attention tracking, emphasizing the necessity of combining multiple sensing methods to achieve more accurate, granular, and effective attention-aware systems.
\par Among these multimodal approaches, eye-tracking has emerged as a particularly valuable method due to its ability to capture subtle visual patterns associated with attentional states. Eye-tracking technology has increasingly gained attention due to its ability to capture subtle visual attention patterns. For instance, Annerer et al. \cite{annerer2021reliably} explored the extent to which eye-tracking metrics reliably distinguish between internally and externally directed attention during numerical, verbal, and visuo-spatial tasks. Their findings indicated that certain eye parameters consistently correlate with an internal focus of attention. Building upon these insights into eye-tracking indicators of attentional states, Wisiecka et al. \cite{wisiecka2022dynamics} utilized webcam-based eye tracking to monitor students' attention distribution during online lectures. Their findings emphasize the importance of managing distractions, optimizing visual layouts, and leveraging real-time attention monitoring to improve learning outcomes.
\par Recognizing the complementary strengths of EEG and eye-tracking, several studies have started integrating these modalities to achieve richer and more reliable attention monitoring \cite{kulke2016neural}\cite{dimigen2011coregistration}\cite{vortmann2019real}\cite{vortmann2022multimodal}.  For example, \cite{vortmann2019real} implemented a system capable of classifying internal versus external attention in real time with over 60\% accuracy, demonstrating that integrating EEG with eye-tracking outperforms single-modality approaches. However, this system fails to classify more attention states. In their subsequent study, Vortmann et al. \cite{vortmann2022multimodal} systematically investigated optimal EEG and eye-tracking integration methods for distinguishing attentional directions, concluding that effective modality fusion depends on specific data characteristics and the intended application context. Collectively, these studies illustrate a clear trajectory toward integrating multimodal sensing approaches, particularly EEG and eye-tracking, to enable more precise, adaptive, and effective attention-aware systems.

\par Most studies focus on detecting and measuring user attention and classifying attention levels. These attention-aware systems restrict the measurement of attention as categories instead of measuring attention levels as continuous values, and they failed to combine eye tracking and EEG to obtain more accurate attention levels. They also do not adjust interfaces according to different attention levels. This gap led to the urgency of developing real attention-aware LLMs to track user attention in real time and dynamically adjust the response and interaction mode. There are three research questions inferred from the gap:
\par 1.How can EEG and eye‑tracking signals be combined to classify a user’s attention into five states (High, Stable, Dropping, Cognitive Overload, Distraction) in real time?
\par 2. How should prompt‑engineering templates be designed so that an LLM dynamically adjusts content depth, length, and interaction style in response to each classified attention state?
\par 3. To what extent will real‑time attention‑aware adaptation enhance user engagement, task performance, and reduce perceived cognitive load compared to a static LLM interface?
\par With the rapid advancements in wearable EEG and eye-tracking technology, and the ubiquity of powerful LLMs, we are now at a point where integrating these components is feasible. This project is timely because it leverages state-of-the-art AI (LLMs) and physiological sensing to tackle the long-standing issue of user disengagement. This study aims to propose an attention aware LLM which will integrate EEG, eye-tracking and real-time data processing tools to detect and create a sliding window that updates dynamically based on real-time EEG+eye-tracking data. An LLM user study will be carried out with this LLM and interaction data will be collected and analyzed. 

\section{Methodology}
\subsection{Participant Recruitment and Task Design}
We will recruit ten healthy university students, all of whom are native English speakers from diverse academic disciplines.  Each participant will complete five brief cognitive task blocks, each lasting between one and three minutes and separated by thirty-second rest intervals, with the order of blocks randomized across participants. The first block will induce high attention by asking participants to read a complex technical passage and then answer comprehension questions. In the second block, stable attention will be maintained as participants solve moderate-difficulty arithmetic problems at their own pace. The third block will simulate dropping attention through the repetitive transcription of a fixed sentence or digit sequence. To provoke cognitive overload in the fourth block, participants will perform timed mental arithmetic while memorizing lists of words. Finally, the fifth block will create distraction by having participants read a neutral text while intermittent pop-up notifications and brief auditory beeps occur. By scripting each block to correspond to one of the five predefined attentional states and by synchronizing continuous EEG and eye-tracking recordings via the Lab Streaming Layer, every overlapping five-second segment of data will automatically inherit its block’s label. 
\subsection{Self‑Report Thought Probes and Labeling Strategy}
To improve the validity of task-based labels, brief thought probes will be presented at pseudo-random intervals ranging from 30 to 60 seconds. Each probe will remain on screen for no more than three seconds and ask participants to rate their attentional state during the preceding period on a five-point scale, with 1 indicating complete distraction and 5 indicating full focus. Responses will be recorded via keyboard or touchscreen and time-aligned with the ongoing physiological data stream. After the session, two trained expert raters will review synchronized video and signal recordings in ELAN (EUDICO Linguistic Annotator) \cite{wittenburg2006elan}, confirm or correct each five-second segment’s label, and resolve any disagreements by consensus. This two-step verification will secure high-quality ground-truth annotations for classifier training.
\subsection{Multimodal Signal Acquisition and Synchronization}
During each task block, we will record electroencephalography (EEG) at 250 Hz using a lightweight wearable headset and capture eye‐tracking data at 60 Hz, including gaze coordinates, pupil diameter, saccades and blinks. To guarantee precise temporal alignment among the EEG stream, the eye‐tracking stream, task markers and thought‐probe events, we will timestamp and merge all data via the Lab Streaming Layer (LSL), achieving sub-millisecond synchronization. 
\subsection{Multimodal Data Preprocessing and Sliding-Window Segmentation}
The continuous EEG data will be filtered using linear finite impulse response filters, applying a high-pass filter at 0.25 Hz and a low-pass filter at 48 Hz. To correct for EOG-related artifacts (electrooculographic signals caused by eye movements and blinks), independent component analysis (ICA) will be performed. Independent components (ICs) identified as corresponding to EOG artifacts will then be removed, following established procedures \cite{delorme2007enhanced}. These components will be identified based on their characteristic spatial and temporal patterns, such as prominent activity over frontal electrode sites and sharp transient waveforms that are commonly associated with blinks. 
For the eye-tracking data, the X- and Y-coordinates as well as pupil diameter measurements from both the left and right eye will be screened for validity. Any samples containing non-existent or missing values will be discarded to ensure data integrity. In addition, binocular blink events, which are automatically identified by the eye-tracker’s built-in detection algorithm, will be excluded from the dataset. The retained X- and Y-coordinate values will be interpreted as the participant’s current gaze location relative to the display surface, forming the basis for downstream analysis such as fixation behavior and gaze variability. The resulting cleaned EEG and eye-tracking streams will feed into two parallel circular buffers, each maintaining the most recent five seconds of data in real time. Every second, the system will append the newest data and drop the oldest, extracting an overlapping five-second segment that remains precisely aligned across both modalities and carries the task-induced attentional label.
\subsection{Feature Extraction}
Power spectral density in the theta (4–7 Hz), alpha (8–12 Hz), and beta (13–30 Hz) frequency bands will be estimated from the EEG data, most likely using the fast Fourier transform (FFT) method. These frequency-specific power values will then be used to compute an engagement index, defined as the ratio of beta power to the combined power of alpha and theta bands ($\beta / (\alpha + \theta)$), which has been widely used as a proxy for cognitive attention or mental effort. In parallel, the eye-tracking data will be used to derive a set of features characterizing visual attention. These will include the average duration of fixations, the spatial dispersion of gaze points (as a measure of scanning behavior), the frequency of saccades (per second), blink rate, and the variability of pupil diameter over time. Once extracted, all EEG and eye-tracking features will be combined into a single vector, typically comprising nine to ten dimensions, to represent the user’s attentional state during each analysis window.
\subsection{Feature Fusion and Attention Classification}
We will combine the EEG and eye-tracking features into a single multimodal vector and input it into a multilayer perceptron classifier. The model will consist of an input layer matching the feature vector, one hidden layer, and a softmax output layer with five units corresponding to the target attentional states. We will train the classifier on our labeled pilot data using cross-entropy loss and will apply early stopping to prevent overfitting. Once deployed, the classifier will process each incoming five-second window and will assign it to one of the five states—High Attention, Stable Attention, Dropping Attention, Cognitive Overload or Distraction—in under one hundred milliseconds, enabling near real-time tracking of user attention.
\subsection{LLM Prompt Adaptation \& Adaptive LLM Output}
In the adaptive condition, attentional state labels classified from real-time physiological signals will be used to dynamically adjust the large language model’s (LLM) responses. For each incoming data window, the system will map the detected attentional state to a corresponding system-level prompt template, which defines the appropriate response style, content depth, and interaction strategy.
The following section details how the system’s response style and interface feedback are modulated based on the five classified attentional states. 
\subsubsection*{High Attention State}
\begin{itemize}
  \item \textbf{Interaction Style}: Detailed explanations, guided reasoning, and in-depth exploration.
  \item \textbf{Information Structure}: Long-form, technical responses including multi-step reasoning or interdisciplinary content. According to \cite{csikszentmihalyi1990flow}, When users are in a high-attention state where their skill level matches the task demands, they are more likely to enter a state of flow. Providing deeper, more complex content is key to sustaining this immersive experience.
  \item \textbf{Visual Feedback}: A clean, distraction-free interface to promote immersion (e.g., Focus Mode).
  \item \textbf{Engagement Strategy}: Offers optional deep-dive prompts such as ``Read More'' or ``Explore Further''.
\end{itemize}

\subsubsection*{Stable Attention State}
\begin{itemize}
  \item \textbf{Interaction Style}: Maintains a smooth, steady conversational tone.
  \item \textbf{Information Structure}: Balanced between depth and simplicity; uses bullet points when helpful.
  \item \textbf{Visual Feedback}: Retains the default interface, allowing minor user personalization.
  \item \textbf{Engagement Strategy}: Encourages light interaction, such as confirming understanding or posing a low-effort question.
\end{itemize}
\subsubsection*{Dropping Attention State}
\begin{itemize}
  \item \textbf{Interaction Style}: Shifts to interactive formats such as quick Q\&A or light gamified prompts.
  \item \textbf{Information Structure}: Shortens paragraphs, highlights key points, and reduces complexity.
  \item \textbf{Visual Feedback}: Introduces subtle UI cues (e.g., color highlights or font emphasis) to redirect focus.
  \item \textbf{Engagement Strategy}: Injects curiosity via humor or unexpected questions to re-engage the user.
\end{itemize}

\subsubsection*{Cognitive Overload}
\begin{itemize}
  \item \textbf{Interaction Style}: Minimizes cognitive demands with simplified, digestible summaries.
  \item \textbf{Information Structure}: Uses concise bullet points and step-by-step instructions instead of dense blocks.
  \item \textbf{Visual Feedback}: Softens UI, hides non-essential elements, and enables pause or review.
  \item \textbf{Engagement Strategy}: Encourages users to take breaks or access “Key Points Summary” prompts to mitigate mental fatigue. 
\end{itemize}

\subsubsection*{Distraction State}
\begin{itemize}
  \item \textbf{Interaction Style}: Redirects attention with short, direct prompts; avoids open-ended or abstract content.
  \item \textbf{Information Structure}: Presents a single clear message at a time, framed as a question or bold takeaway to quickly capture interest.
  \item \textbf{Visual Feedback}: Introduces animated or attention-catching UI elements (e.g., pulsing icons, micro-interactions) to draw the user’s gaze.
  \item \textbf{Engagement Strategy}: Offers curiosity hooks or task-relevant surprises (e.g., “Did you know?” facts) to reorient user focus and reduce disengagement.
\end{itemize}

\subsection{User Study and System Evaluation}
\subsubsection{Participants \& Experimental Design}
We will recruit twenty healthy university students aged 18–35, all of whom are native English speakers from diverse academic disciplines. Participants will be screened to ensure normal or corrected-to-normal vision. Written informed consent will be obtained from all participants prior to their participation, following the ethical standards approved by the institution’s IRB. 
A within-subjects, counterbalanced design will be used, with session order randomized to control for learning and order effects.
\subsubsection{Materials and Equipment}
In the adaptive condition, participants will interact with attention-aware LLM that integrates real-time electroencephalography (EEG) and eye-tracking signals to dynamically adjust the depth of content, interaction style, and visual presentation cues based on inferred attentional states. In contrast, during the baseline condition, participants will engage in a standard LLM in which system responses remain invariant and are not modulated by physiological signals, regardless of fluctuations in user attention.

Task stimuli will encompass four types of activities: decision making, creative problem solving, debating, and information seeking. For decision making tasks, participants will be presented with a complex real-world scenario that requires weighing multiple factors before making a decision. The scenario will outline a situation with competing options, each associated with trade-offs and risks. Participants will be instructed to discuss the options in detail, evaluating advantages and disadvantages, and progressively refine their decision-making process through structured dialogue. They will be encouraged to explore alternative perspectives, reconsider initial choices based on new information, and justify their final decision. For creative problem solving tasks, participants will be given an open-ended challenge that requires generating innovative solutions to a practical or conceptual problem. The initial prompt will define a broad goal or constraint, but will not prescribe specific strategies. Participants will be instructed to brainstorm ideas, elaborate on partial solutions, combine different approaches, and refine their proposals iteratively. They will be encouraged to critically evaluate the feasibility of their ideas, explore modifications, and synthesize multiple contributions into coherent final solutions. For debating tasks, participants will be presented with a controversial issue involving two opposing viewpoints. They will first be instructed to engage with one perspective, constructing arguments and gathering supporting points. Subsequently, they will be asked to switch perspectives and develop counterarguments. Throughout the process, participants are encouraged to challenge assumptions, seek clarifications, request elaborations on specific points, and refine the arguments through iterative exchange. For information seeking tasks, participants will be given a multi-part factual inquiry that requires gathering, organizing, and integrating information from multiple dimensions. They will be instructed to iteratively seek clarifications, request elaborations, and gradually build a structured response to the inquiry. All task materials will be counterbalanced across conditions to ensure that each participant encounters equivalent content in both adaptive and baseline sessions.

Physiological data will be collected using a single-channel wearable EEG headset for neural activity monitoring and a screen-mounted infrared eye-tracking system to capture gaze behavior and pupil dynamics. Participants will complete all interactions using a standard desktop personal computer equipped with a full-size keyboard and optical mouse for input.
\subsubsection{Procedure}
The experiment will occur in a single laboratory visit lasting approximately ninety minutes. Upon arrival, participants will be welcomed and seated approximately 60 centimeters from a desktop monitor. They will be fitted with a single-channel EEG headset and complete a calibration procedure for the screen-mounted eye-tracker. After verifying signal quality across both EEG and eye-tracking devices, participants will complete a brief five-minute familiarization phase. During this phase, participants will complete a brief practice interaction with the attention-aware language model, during which they can enter a sample query and observe how the system responds. This familiarization phase is not included in the data analysis.
Following familiarization, each participant will take part in two experimental sessions: one involving the adaptive interface and the other using a non-adaptive baseline system. The order of these sessions will be randomized. Each session will consist of four distinct five-minute tasks: decision making, creative problem solving, debating, and information seeking. Each task will be separated by a thirty-second rest interval during which participants can relax but must remain seated. Upon completing both sessions, participants will complete a post-task battery consisting of the NASA Task Load Index (NASA-TLX), which will be used to evaluate participants’ subjective cognitive workload across standard dimensions such as mental demand, effort, and frustration. Finally, participants will engage in a brief semi-structured interview lasting approximately ten minutes, during which they will provide qualitative feedback regarding their experiences, perceived benefits, and potential drawbacks of adaptive versus baseline interactions. 

\subsubsection{Data Collection}
Electroencephalography (EEG) and eye-tracking signals will be recorded throughout all task periods. EEG signals will be sampled at 250 Hz using a single-channel wearable headset, while eye-tracking data—including gaze position, fixation duration, blink events, and pupil diameter—will be captured at 60 Hz via an integrated screen-mounted infrared tracker. All streams will be time-synchronized via the Lab Streaming Layer (LSL) protocol to ensure sub-millisecond alignment with task event markers. Detailed behavioral interaction logs will be collected, comprising system timestamps of participant queries and responses, task start and end times, number of follow-up prompts requested, and frequency of clarification or elaboration queries issued during each session. Subjective assessments will be collected following the completion of both sessions. 
\subsubsection{Data Analysis}
We will capture four types of outcome measures. First, user engagement will be quantified through objective behavioral metrics, including time on task, number of follow-up prompts requested, and gaze metrics such as fixation count and fixation duration, derived from the eye-tracking data stream. Second, task performance will be assessed where applicable, based on response accuracy and response time in structured tasks such as decision making. For open-ended tasks where accuracy is not directly defined, performance will be evaluated through qualitative ratings of output quality across dimensions such as coherence, creativity, and persuasiveness. Third, subjective cognitive load will be measured immediately after each session using the NASA Task Load Index (NASA-TLX), capturing both overall workload and subscale scores. Fourth, physiological engagement will be assessed through EEG-derived metrics, specifically the engagement index.

Quantitative analyses will compare participant outcomes between the adaptive and baseline conditions using repeated-measures ANOVA. When assumptions of sphericity are violated, Greenhouse-Geisser corrections will be applied to adjust the degrees of freedom. If normality assumptions are not met, non-parametric alternatives such as the Wilcoxon signed-rank test will be used instead. Effect sizes will be reported alongside corresponding p-values to aid interpretation. In addition to these inferential analyses, exploratory correlations will be conducted to investigate associations between physiological engagement indices, behavioral interaction patterns, and subjective workload ratings.

In addition to quantitative analysis, qualitative data will be collected through brief post-task interviews focused on perceived clarity, helpfulness, adaptivity, and mental effort during the interaction. Interviews will be transcribed verbatim and analyzed using inductive thematic analysis. Two independent raters will iteratively generate, refine, and reconcile codes, resolving any discrepancies through discussion to ensure coding reliability.

\section{Milestones \& Project Plan}

Phase 1: Literature Review \& System Design (Weeks 1-4) Review existing research on attention detection, EEG-based cognition models, and adaptive interfaces.
 
Design the system architecture integrating EEG, eye-tracking, and LLM components.  
Plan the experimental tasks and prepare study protocols (IRB considerations, consent forms if needed).

Phase 2: Data Collection \& Model Development (Weeks 5-8) Recruit participants and run EEG + eye-tracking experiments.
  
Preprocess and label the data with attention state annotations.  
Develop the attention classification model; extract features and train the model, iterating to improve accuracy.

Phase 3: LLM Integration \& Adaptive Response Mechanism (Weeks 9-12) Implement real-time LLM adaptation based on detected attention states.

Connect the trained attention classifier to trigger adaptive responses in the LLM.  
Integrate UI/UX adjustments (e.g., dynamic text highlighting, interface theme changes) corresponding to attention states.  
Conduct preliminary internal tests to ensure system stability and low-latency performance.

Phase 4: User Testing \& System Evaluation (Weeks 13-16) Full user study with attention-aware LLM.

Conduct a user study with the attention-aware LLM (using a subset of participants, e.g., 10-15, as a pilot). Collect user feedback, engagement metrics, and cognitive load assessments during these sessions.  
Analyze the results to evaluate the impact of attention-aware adaptation. Refine the model or system based on findings.  
Finalize the project report and presentation, including lessons learned and future improvement suggestions.

\section{Expected Results}
We expect that integrating EEG and eye-tracking will enable reliable detection of user attention states in real time. Specifically, we anticipate achieving at least 70\% accuracy in classifying the five defined attention states using our model. Furthermore, we hypothesize that users interact with the attention-aware LLM will demonstrate improved engagement and reduced cognitive load compared to interactions with a standard LLM. Users are expected to maintain attention for longer durations, exhibit better recall of provided information, and report lower levels of frustration or mental fatigue. Additionally, we anticipate observing distinct physiological signatures (EEG and eye-tracking) for each attentional state, enabling us to clearly differentiate among the predefined states.

Qualitative feedback will further confirm whether adaptive responses enhance user satisfaction and perceived system responsiveness. We also expect to identify optimal EEG and eye-tracking features and determine the most effective machine learning algorithms for real-time attentional state classification. If successful, this project will result in a working prototype supported by preliminary user study data demonstrating the effectiveness of adaptive LLM responses in enhancing overall user experience and validating the feasibility of neuroadaptive technologies in interactive AI systems.

Finally, by examining feature importance scores and comparing alternative classifiers (such as random forests or lightweight convolutional networks), we will identify the most informative EEG and eye-tracking features and the most effective algorithmic approach for low‐latency attention detection. Together, these results will validate the feasibility and benefits of real-time neuroadaptive LLM interactions and guide future refinements of attention-aware AI systems.
\section{Risks \& Mitigation Strategies}

\par Risk 1: Noisy or low-quality EEG/eye-tracking data. Mitigation: Apply robust preprocessing (filtering, artifact removal) and calibration for each user; possibly discard segments of data with excessive noise. Collect sufficient training data so that the model can generalize despite noise.
\par Risk 2: Classification accuracy may be low for subtle attention differences. Mitigation: Use state-of-the-art machine learning techniques, such as ensemble classifiers or deep learning, and consider fusing EEG and eye features to improve robustness. Validate the model on diverse data (maybe use cross-validation or augment with existing datasets if available) to ensure it generalizes.
\par Risk 3: High latency in real-time processing affecting user experience. Mitigation: Optimize the system pipeline – for example, use lightweight models for real-time prediction, and ensure efficient data streaming from sensors to the computer. We may also window the data to balance responsiveness with accuracy. If needed, some adaptation features might be simplified to reduce computation. Careful synchronization of the components (EEG device, eye-tracker, and LLM interface) is planned to minimize lag.
\par Risk 4: Integration complexity (multiple components could fail). Mitigation: We will modularize the system so that failure in one component (e.g., a glitch in the EEG feed) does not crash the whole system. For instance, if EEG data is momentarily unavailable, the system could fall back on eye-tracking data or pause adaptation. We will test each module independently (unit testing) and then together to ensure reliability.
\par Risk 5: User privacy and comfort concerns. Mitigation: As you wrote, implement strict data privacy protocols: encrypt stored physiological data, anonymize it for analysis, and obtain informed consent with full transparency about what data is collected and how it will be used. Additionally, to address user comfort, we will emphasize that the system is there to help and that users can opt out at any time. All experiments will be approved by an ethics review board (if applicable for the course) and conducted with care for the participants’ well-being.
\par Risk 6: Participants may exhibit substantial individual differences in EEG and eye-tracking signals, potentially reducing the accuracy and generalizability of our model across diverse users. Mitigation: We will use cross-validation techniques and participant normalization methods to enhance model robustness. Additionally, training our models on diverse participant data and implementing personalized calibration sessions prior to main experiments can further improve individual-level accuracy and overall model performance.
\newpage

\bibliographystyle{plain} 
\bibliography{ref} 

\end{document}